\documentclass[%
reprint,
superscriptaddress,
table,
 amsmath,amssymb,
 aps,
floatfix,
]{revtex4-2}

\usepackage{graphicx}% Include figure files
\usepackage{dcolumn}% Align table columns on decimal point
\usepackage{bm}% bold math
\usepackage{braket}
\usepackage[colorlinks=true, linkcolor=blue]{hyperref}
\usepackage[most]{tcolorbox}
\definecolor{boxcolor}{RGB}{239,239,235}
\usepackage{wrapfig}
\usepackage[capitalize]{cleveref}
\usepackage{multirow}
\usepackage{makecell}
\usepackage[position=top]{subfig}
\usepackage{threeparttable} % to use table notes
\usepackage{upgreek}

\renewcommand{\arraystretch}{2.5}
\newcolumntype{s}{>{\columncolor[HTML]{AAACED}} p{3cm}}

\begin{document}

\preprint{APS/123-QED}

\title{New opportunities in condensed matter physics for nanoscale quantum sensors} % Force line breaks with \\
\author{Jared Rovny} 
\affiliation{Princeton University, Department of Electrical and Computer Engineering, Princeton, NJ 08544, USA}
\author{Sarang Gopalakrishnan} 
\affiliation{Princeton University, Department of Electrical and Computer Engineering, Princeton, NJ 08544, USA}
\author{Ania C. Bleszynski Jayich} 
\affiliation{Department of Physics, University of California Santa Barbara, Santa Barbara, CA 93106, USA}
\author{Patrick Maletinsky} 
\affiliation{Department of Physics, University of Basel, 4056 Basel, Switzerland}
\author{Eugene Demler} 
\affiliation{Institute for Theoretical Physics, ETH Zurich, 8093 Zurich, Switzerland}
\author{Nathalie P.\ de Leon}
\thanks{Corresponding author. Email: npdeleon@princeton.edu}
\affiliation{Princeton University, Department of Electrical and Computer Engineering, Princeton, NJ 08544, USA}

\date{\today}

\begin{abstract}
Nitrogen vacancy (NV) centre quantum sensors provide unique opportunities in studying condensed matter systems: they are quantitative, noninvasive, physically robust, offer nanoscale resolution, and may be used across a wide range of temperatures. These properties have been exploited in recent years to obtain nanoscale resolution measurements of static magnetic fields arising from spin order and current flow in condensed matter systems. 
Compared with other nanoscale magnetic-field sensors, NV centres have the unique advantage that they can probe quantities that go beyond average magnetic fields.
Leveraging techniques from magnetic resonance, NV centres can perform high precision noise sensing, and have given access to diverse systems, such as fluctuating electrical currents in simple metals \cite{Kolkowitz2015a,Ariyaratne2018a} and graphene \cite{Andersen2019a}, as well as magnetic dynamics in yttrium iron garnet \cite{Du2017a}. In this review we summarise unique opportunities in condensed matter sensing by focusing on the connections between specific NV measurements and previously established physical characteristics that are more readily understood in the condensed matter community, such as correlation functions and order parameters that are inaccessible by other techniques, and we describe the technical frontier enabled by NV centre sensing.

\end{abstract}

\maketitle
  
\section{\label{sec:intro} Introduction}

Much of what we know about condensed matter systems---their structure, excitations, and dynamics---is inferred from bulk transport and spectroscopy. Bulk d.c. transport and optical response characterize the spatially averaged (i.e., zero-momentum) response of a sample when it is globally perturbed. Thus they only offer an indirect probe of theoretically fundamental properties such as how excitations disperse.  Spectroscopic techniques that do offer momentum resolution, such as neutron scattering, often require large sample volumes to achieve measurable signal, making it challenging to study low-dimensional materials. 
Moreover, the spatially averaged data that bulk probes---as well as surface probes like ARPES---give can be incomplete or even misleading in systems whose response is essentially spatially heterogeneous: the average can be dominated by atypical parts of the sample~\cite{vojta2010quantum}. Heterogeneity can arise because the sample is disordered or because interactions give it a local tendency to phase-separate~\cite{PhysRevLett.94.056805}, inducing the spontaneous formation of static or fluctuating domains~\cite{huang2017numerical}. Thus, quantitatively understanding such phenomena requires local, momentum-resolved probes.

There are several established nanoscale probes: notably, scanning-tunneling microscopy (STM) measures the local electronic density of states \cite{Yazdani2016a,Yin2021a,Jack2021a}, scanning SQUID microscopy measures local magnetic fields \cite{Persky2022a,Marchiori2022a}, and microwave impedance microscopy (MIM) measures local conductivity and permittivity \cite{Barber2021a}. While these probes have led to immense advances in our understanding of materials, electronic structure, and correlated phases, their modes of operation can restrict the available parameter space for the material under investigation (for example to low temperatures), and it is difficult to gain access to local magnetic field textures and dynamics with these probes. 
In the past decade, nitrogen-vacancy (NV) centres in diamond have emerged as a new class of nanoscale sensor that complements these technologies in their temperature range and in their ability to probe both static and dynamic properties in a momentum- and frequency-resolved way.

In essence, an NV centre is an optically addressable qubit that can sense its environment with extremely high spatial resolution \cite{Maze2008a,Abobeih2019a,Marchiori2022a}. The spatial resolution is determined by the offset between the NV centre and the material of interest (typically $5$ to $100$ nm), ultimately limited by the extent of the electronic wavefunction (around $1$ nm). Because the NV centre offers both local resolution and the entire suite of qubit-manipulation techniques developed in the context of nuclear magnetic resonance~\cite{vandersypen}, it enables a host of new local sensing technologies. As an example of how versatile these probes are, static magnetic fields shift the qubit splitting \cite{Maze2008a,Balasubramanian2009a}, the noise spectral density at the qubit transition frequency can change the qubit's $T_1$ time \cite{Kolkowitz2015a,Ariyaratne2018a}, and dynamic structure like the spatio-temporal correlations of the noise in an underlying sample can be detected in a coherent ($T_2$ limited) measurement of multiple NV centres \cite{Rovny2022a}. 

The momentum resolution of an NV centre comes from changing its distance from the sample (or using an ensemble of NV centres at different distances). Notably, the momentum-resolved current noise that NV centres measure goes beyond what one can measure using techniques like neutron scattering. Probes like neutron scattering \cite{Boothroyd2020}, X-ray scattering \cite{Ament2011}, and electron energy loss spectroscopy \cite{Vig2017} measure the momentum-dependent \emph{density} response, from which one can extract the behavior of longitudinal currents (i.e., those along the wavevector being probed) but not transverse ones. The current noise that an NV centre detects can also be sensitive to transverse current fluctuations, and can therefore pick up the response of the electron fluid to being sheared; the shear response is a key diagnostic of Wigner crystallization \cite{Dolgirev2023}, as well as a natural probe of viscous hydrodynamic behavior \cite{Agarwal2017,Lucas2018}. Furthermore, while neutron scattering averages over the bulk of the material, NV centres enable real-space probing with nanoscale resolution, giving access to spatially inhomogeneous momentum information.

\begin{figure}[ht]
	\includegraphics[width=0.44\textwidth]{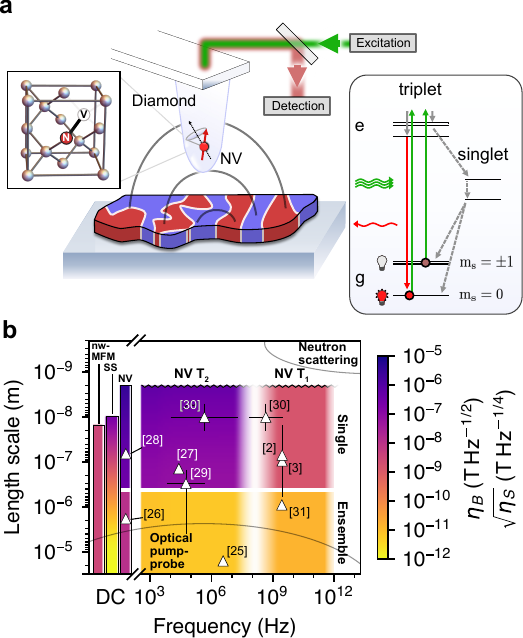}
	\caption{(a) Overview of an NV centre experiment using a diamond scanning tip. An NV centre in diamond detects local fields from a target system with nanometer resolution. Above-bandgap excitation light is used to spin-polarize the NV centre, and subsequently read out its state through spin-dependent fluorescence. (b) Approximate frequency ranges and length scales accessible through common NV centre sensing modalities (see Table 1 for details and Box 1 for methods bridging the $T_2$, $T_1$ gap). Optimal sensitivities assume either the minimum detectable field using coherent $T_2$ sensing ($\eta_B$), or the minimum detectable mean square noise in a 1 Hz band using incoherent $T_1$ relaxometry ($\eta_S$). Selected NV centre experiments are indicated \cite{Ariyaratne2018a,Andersen2019a,Glenn2018a,Hart2021a,Vool2021a,Woernle2021a,Monge2023a,Sangtawesin2019a,Huang2023}, where achieved sensitivities vary. At DC, comparisons are also made with nanowire magnetic force microscopy (nw-MFM) and scanning SQUID (SS) measurements \cite{Marchiori2022a}. Neutron scattering and optical pump/probe experiments show approximate length- and time-scales of system excitations, while for NV centres the length-scales refer to the NV-sample distance or the square root of areal sensing resolution for ensembles. Note that this plot is simplified, as e.g.\ shallow ensembles and $T_2$ noise spectroscopy are also used (see text).}
	\label{fig:overview}
\end{figure}

Thus, NV centres offer a new suite of tools for condensed matter experiment---including many tools that are only just being developed. NV centres have been employed to measure static local fields from antiferromagnetic order \cite{Appel2019a,Finco2021a}, multiferroic \cite{Haykal2020a} order, and from currents in materials with electron hydrodynamics \cite{Vool2021a,Jenkins2022a} and superconductivity \cite{Thiel2016a,Pelliccione2016a,Schlussel2018a,Monge2023a}, as well as linear dynamics in contexts such as magnon imaging \cite{Lee-Wong2020,Simon2022a} and conductivity imaging \cite{Ariyaratne2018a}. The last few years have also seen exciting steps exploring out-of-equilibrium systems, such as a magnon scattering platform \cite{Zhou2021a} and measurements of an electron-phonon instability in graphene \cite{Andersen2019a}, where newly developed NV centre sensing methods like covariance magnetometry \cite{Rovny2022a} and gaining access to the noise correlation time via coherence decay \cite{Dwyer2022a,Davis2023a,Joos2022} may provide insights that are difficult to access with other sensing techniques.

The present review aims to outline the physics of NV centres as it relates to condensed-matter sensing experiments, and to illustrate a few recent developments and proposals in this area. We focus on the parametrization of different NV centre sensing modalities in terms of physical quantities of interest for condensed matter physics and review the current state of the art, including new sensing techniques and sophisticated methods in image reconstruction, efficient sampling, and sensitivity enhancement that have been developed within the last few years. We conclude with a discussion of new opportunities in condensed matter sensing that are enabled by this unique probe and toolkit.

\section{The physics of NV centres}
\label{Sec:Techniques}

The NV centre is a point defect in diamond consisting of a substitutional nitrogen next to a vacancy in the lattice. Its ground state electronic configuration has spin $S = 1$, and this ground state spin can be initialized and measured even at room temperature, with spin coherence times up to a few milliseconds \cite{Balasubramanian2009a}, and at cryogenic temperatures with coherence times exceeding 1 second \cite{Abobeih2018a}. The long spin coherence time is enabled by weak spin orbit coupling, low electron-phonon coupling, and a quiet spin environment in ultrapure diamond \cite{Childress2013a}. It can be used as a sensor over a remarkably wide range of parameters, from cryogenic temperatures to nearly 1000 K \cite{Liu2019a}, and from vacuum to GPa pressures \cite{Hsieh2019a,Lesik2019a}, largely because its electronic level structure allows for nonresonant optical initialization and readout, and because its structure in diamond is exceptionally robust. Typically, NV centres are deployed in three broad categories of platforms (see Box 1) enabled in part by sophisticated diamond fabrication techniques. These platforms range from cantilevers incorporating individual embedded NV centres with nanometer scanning resolution to large 2D ensembles of $\sim 10^{12}$ NV centres per cm$^2$ simultaneously imaged across millimeters. Each platform has inherent tradeoffs, detailed in Box 1, including spatial resolution versus sensing time (scanning tips are higher-resolution but slow, while widefield imaging achieves faster measurements by ensemble averaging at the expense of spatial resolution) and spatial resolution versus coherence time (NV centres exhibit shorter coherence times as they approach the diamond surface). In many cases, a system of interest may be measured using multiple modalities to paint a more complete picture of its properties \cite{Ku2020a}.

\begin{table*}[t] % top of page
\begin{threeparttable}
\setlength\tabcolsep{5pt}
\renewcommand{\arraystretch}{2.5}
  \centering
  \small
  \subfloat{%
  \begin{tabular}{cc|cc|cc} 
 \rowcolor{boxcolor}[\tabcolsep][1.5em] \multicolumn{2}{c}{} & \multicolumn{2}{c}{\textbf{Resolvable NV centres}} & \multicolumn{2}{c}{\textbf{Ensemble NV centres}} \\
 \rowcolor{boxcolor}[\tabcolsep][1.5em] \multicolumn{2}{c|}{} & Deep & Shallow ($\sim 5$ nm) & Deep & Shallow ($\sim 5$ nm) \\
 \hline
\multicolumn{2}{c|}{T$_2^*$ ($\mu$s)} & 470\tnote{*}\, \cite{Maurer2013a} & 
- & 
\makecell{
0.001 ppm: 8.7\tnote{*}\, \cite{Barry2023a} \\ 
10 ppm: 1\tnote{*}\, \cite{Bauch2020a}} & 
- \\
\multicolumn{2}{c|}{T$_2$ ($\mu$s)} & 
1800\tnote{*}\,  \cite{Balasubramanian2009a} & 
43 \cite{Sangtawesin2019a} & 
\makecell{
0.3 ppm: $200$ \cite{Wang2021a} \\ 
10 ppm: 20\tnote{*}\, \cite{Bauch2020a}} & 
$\sim$100/$\upmu$m$^2$: 45\tnote{*}\, \cite{Rodgers2023a} \\
\multicolumn{2}{c|}{T$_1$ (ms)} & 
7.5\tnote{*}\, \cite{Maurer2013a} & 
2 \cite{Sangtawesin2019a} & 
\makecell{
3.3 ppm: 2.5 \cite{Jarmola2015a} \\ 
45 ppm: 0.067 \cite{Choi2017a}} & 
$\sim$100/$\upmu$m$^{2}$: 0.75 \cite{Liu2022a}, 5\tnote{*}\, \cite{Rodgers2023a} \\
\hline
\rowcolor{boxcolor}[\tabcolsep][1.5em] \multicolumn{2}{c|}{} &\multicolumn{2}{c|}{$\text{unit}/\sqrt{\text{Hz}}$} & \multicolumn{2}{c}{$\text{unit}/\sqrt{\text{Hz}\,\upmu\text{m}^3}$} \\ 
\rowcolor{boxcolor}[\tabcolsep][1.5em] \multicolumn{2}{c|}{\multirow{-2}{*}{\textbf{Sensitivity}}} & Deep & Shallow ($\lesssim 100$ nm) & Deep & Shallow ($\lesssim 100$ nm)  \\
\multirow{2}{*}{\makecell{Magnetic \\ ($\text{nT}$)}}  & DC & 
72\tnote{*,$\dag$}\,\hspace{1.5mm} (bd) \cite{Balasubramanian2009a} & 
\makecell{
300 (st) \cite{Sun2021a} \\
4300 (st) \cite{Woernle2021a}} & 
\makecell{
24.6\tnote{*}\, (qdm) \cite{Tang2023} \\
11.5 (bd)\cite{Barry2023a}}
& \makecell{
176 (bd) \cite{Waxman2014a}, \\
1440 (dac) \cite{Hsieh2019a}} \\
\multirow{2}{*}{} & AC & 4.3\tnote{*}\,  (bd) \cite{Balasubramanian2009a} & 50 (st) \cite{Palm2022a} & \makecell{2.4\tnote{*}\, (bd) \cite{Glenn2018a} \\
5.3 (bd) \cite{Barry2023a}} & \makecell{See \cite{Liu2022a,Rodgers2023a} \\ } \\
\multirow{2}{*}{\makecell{Electric \\ ($\text{mV}/\mu\text{m}$)}} & DC &
89.1 (bd) \cite{Dolde2011a} & 
3520 (st) \cite{Bian2021a} & 
2217 (bd) \cite{Chen2017a} & - \\
\multirow{2}{*}{} & AC & 20.2 (bd) \cite{Dolde2011a} & 24-26 (st) \cite{Qiu2022a,Huxter2023a} & 11\tnote{*}\, (bd) \cite{Michl2019a} & - \\
\multicolumn{2}{c|}{Stress (MPa)} & - & 
0.6\tnote{$\dag$}\, (dac) \cite{Doherty2014a} & 
$0.06$\tnote{*}\, (qdm) \cite{Marshall2022} & 
3.6 (dac) \cite{Hsieh2019a} \\
\multicolumn{2}{c|}{\makecell{Temperature \\ (mK)}} & 
5\tnote{*}\, (bd) \cite{Neumann2013a} & 
24 (bd) \cite{Wang2015a} & 
194 (bd) \cite{Shim2022} & 
\makecell{
36.8 K: 20 (nd) \cite{Hui2019} \\ 
118 K: 82 (nd) \cite{Hui2019} \\
300 K: 4.8 (bd) \cite{Moreva2020} \\
1000 K: 9 (nd) \cite{Liu2019a}}  
  \end{tabular}
  }
    \begin{tablenotes}
       \item [*] Makes use of $^{12}$C isotopically enriched diamond.
       \item [$\dag$] Derived using values from separate measurements \cite{Balasubramanian2009a,Doherty2014a}.
    \end{tablenotes}
  
  \caption{Demonstrated properties of different NV centre sensing platforms, including scanning tip (st), diamond anvil cell (dac), quantum diamond microscope (qdm), nanostructure/membrane (nsm), nanodiamonds (nd), and bulk diamond (bd). The listed $T_2$ are dynamically decoupled. Room temperature measurements are given for $T_1$ and $T_2$. For detailed temperature-dependent NV properties, see \cite{Cambria2023a,Ernst2023,Happacher2023a}. For detailed NV-density dependence, see \cite{Bauch2020a, Jarmola2015a, Choi2017a}. For detailed depth-dependence, see \cite{Sangtawesin2019a,Dwyer2022a,Myers2014}.}
  \label{tab:1}
\end{threeparttable}
\end{table*}

An NV centre measurement typically consists of three steps: initialization by optical polarization into the $\ket{m_s=0}$ spin state, interaction with external fields using a tailored sensing modality (described below and in Box 1), and optical detection through spin-dependent fluorescence (Fig.\ 1a). The three broad sensing modalities, covering characteristic frequency ranges from DC to $\sim$100 GHz, are (A) optically detected magnetic resonance (ODMR) spectroscopy of the $0 \rightarrow \pm1$ transition frequency for DC sensing; (B) coherent Ramsey-based protocols which may incorporate frequency-selective pulse sequences for AC sensing up to $\sim\,$10 MHz frequencies; and (C) relaxometry, which is performed by monitoring the NV centre decay rate $\Gamma$ among the ground state spin levels due to fields fluctuating at the NV transition frequency. The latter technique has been demonstrated up to $\sim\,$100 GHz \cite{Fortman2021a}, where the attainable frequency is determined by the strength of the external magnetic field and the availability of microwave hardware used to drive the NV centre spin transitions. While NV centres are most often used to detect magnetic fields through the induced splitting $\omega = D_{\text{zfs}} + \gamma_eB$, where the zero field splitting is $D_{\text{zfs}}=2\pi \times 2.87\,$GHz and the electron gyromagnetic ratio is $\gamma_e = -2\pi \times28.02\,$GHz/T, they may also be used to detect electric fields, temperature, and strain (Box 1). An important advantage offered by NV centres is the ability to engineer their sensitivity to the environment through pulse sequences designed to select certain frequencies \cite{Degen2017a} or physical observables \cite{Hsieh2019a} (Sec.\ \ref{sec:t2}).

The sensitivity of an NV centre measurement may be boosted by increasing the number of NV centres, extending the coherence time, improving the readout fidelity through techniques like spin-to-charge conversion \cite{Shields2015a} and repetitive readout using nuclear spin ancillae \cite{Jiang2009a,Arunkumar2023a}, and, crucially for nanoscale sensing, decreasing the distance to the target. Typically achievable sensitivities are approximately 1 $\mu\text{T}/\sqrt{\text{Hz}}$ at DC \cite{Zhong2022a} and $50\,\text{nT}/\sqrt{\text{Hz}}$ at AC \cite{Vool2021a,Huxter2022a,Palm2022a} (see Table 1 for details of platform-specific metrics). Spatial resolution down to $\sim\,$20 nm \cite{Chang2017a} has been achieved using scanning tips, while implantation into unstructured substrates allows for NV centres as close as 2-3 nm to the surface \cite{OforiOkai2012a,Romach2015a}, and furthermore, subdiffraction imaging techniques allow the NV centre to be laterally located to within a few nanometers \cite{Gardill2023a,Dolde2013a}. 

While the NV centre spin coherence time and resonance frequency naively imply $\sim\,$kHz linewidth and MHz or GHz frequency ranges for Ramsey and relaxometry, respectively, several protocols have been developed to surpass these limits. For detecting coherent signals, the NV sensor may be reinitialized and detected many times within the coherence time of the signal, enabling sub-millihertz linewidths \cite{Glenn2018a,Schmitt2017a,Boss2017a}. To broaden the accessible frequencies, recent demonstrations have shown that dressed states and frequency mixing may be used to fill in the frequency gap between Ramsey and relaxometry protocols to access $\sim\,$100 MHz frequencies, provided mixing fields with sufficiently high power and frequency are available \cite{Joas2017a,Stark2017a,Meinel2021a,Wang2022a}, while auxiliary spins may be leveraged to access different frequencies while boosting sensitivity \cite{Zhang2023a}. As shown in Fig.\ 2a, new techniques have also been developed to improve spatial resolution and imaging, by reconstructing the source magnetization from the stray field maps measured by the NV centre using machine learning algorithms \cite{Dubois2022a}. 

\onecolumngrid
\begin{tcolorbox}[colback=boxcolor, colframe=boxcolor,title=\textbf{Box 1: NV Centre Modalities and Platforms},coltitle=black,width=\textwidth]

    \vspace*{10pt}
    \begin{center}
        \includegraphics[width=0.95\textwidth]{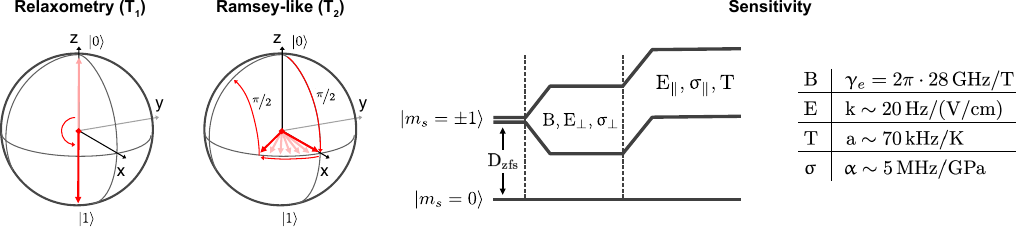}
    \end{center}

    \textbf{Sensing modalities}

    The typical generic NV centre Hamiltonian has the form
    \begin{align}
        \mathcal{H}=\mathcal{A}S_z^2 + \mathcal{B}S_z + \mathcal{C}(S_y^2-S_x^2) + \mathcal{D}(S_xS_y+S_yS_x), \label{eq:Hamiltonian}
    \end{align}
    where (for small applied fields) the separate contributions cause shifting ($\mathcal{A}$) and splitting ($\mathcal{B,C,D}$) of the $\ket{m_s=\pm1}$ states according to the energy diagram above, which describes the approximate couplings to parallel and perpendicular magnetic fields ($\textrm{B}$), electric fields ($\textrm{E}$), stress ($\sigma$), and temperature ($\textrm{T}$); note that these sensitivities may change depending on e.g.\ temperature and field orientation \cite{Toyli2012a,Barson2017a,Huxter2023a}. Continuous wave ODMR provides a simple and robust method for spectroscopy of Hamiltonian \ref{eq:Hamiltonian}, but is insensitive to AC fields and has limited sensitivity to DC fields. Pulsed Ramsey-type protocols employ trains of pulses to modulate the NV centre in the time domain, making the NV centre sensitive to a band of frequencies centered at the pulsing rate. Pulsed methods are extremely versatile, and are routinely used for frequency-selective spectroscopy and to significantly improve measurement sensitivity, including sensitivity to DC fields through lock-in gradiometry \cite{Huxter2022a}. Lastly, incoherent relaxometry, while less versatile, provides access to high-frequency fields inducing single-quantum ($\ket{m_s=\pm1}\rightarrow\ket{m_s=0}$) and double-quantum ($\ket{m_s=\pm 1}\rightarrow\ket{m_s=\mp 1}$) transitions of the NV centre, where this distinction can be useful in diagnosing whether the relaxing fields are electric or magnetic in nature \cite{Myers2017a}. Beyond the simplified Hamiltonian in \cref{eq:Hamiltonian}, it is often advantageous to intentionally mix spin states or probe dressed levels of a driven state \cite{Huxter2023a}. 

    \begin{center}
    \vspace*{10pt}
	\includegraphics[width=0.9\textwidth]{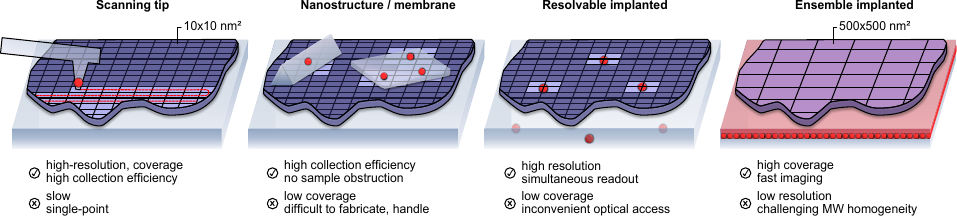}
    \end{center}

    \textbf{Sensing platforms}

    The broad NV centre platform categories are shown above, and include an NV centre embedded in an AFM-style scanning tip, deposited diamond nano- and micro-structures, and low or high densities of NV centres implanted in bulk substrates. Some platform tradeoffs are indicated above, where the most relevant are usually light collection efficiency, spatial resolution, and NV centre coherence. Diamond fabrication and processing have seen significant advances, including high collection efficiency scanning tips \cite{Wan2018a,Hedrich2020a} and ``smart-cut'' nanoscale diamond membranes \cite{Aharonovich2012a,Guo2021a}, where both platforms may be integrated with existing target structures. Close NV-sample proximity, enabled by very shallow ion implantation just a few nanometers from the diamond surface and subsequent fabrication \cite{Behzad2015a,Challier2018a}, incurs both decoherence due to surface noise and charge state instabilities due to surface electronic traps; coherence times have been significantly improved through a better understanding of surface noise sources and treatment \cite{Sangtawesin2019a,Wood2022a,Zvi2023a,Janitz2022a}, although achieving bulk-like properties near the surface remains an open challenge \cite{Rodgers2021a}. Improving sensitivity in a given experiment requires a judicious choice of distance versus coherence time, number of NV centres, and the readout method (which can feature trade-offs between speed and fidelity) \cite{Jiang2009a,Shields2015a,Hopper2018a,Zhang2021a,Irber2021a}. Proximity to the surface is also crucial for spatial resolution, as an NV centre at distance $d$ from the surface is sensitive to stray fields originating from a region on the surface with spatial extent proportional to $d$ (see Box 2). Finally, high-density NV centre ensembles sacrifice spatial resolution for a significant increase in sensitivity; spin-spin interactions place an upper bound on the useful NV centre density, but dynamical decoupling protocols tailored to the spin environment can effectively suppress these interactions \cite{Choi2020a}.
 
\end{tcolorbox}
\twocolumngrid

\section{Static Fields}
\label{Sec:Static}

NV magnetometry offers an unmatched combination of spatial resolution and sensitivity in magnetic imaging\,\cite{Rondin2014a}, together with a wide operational range in, e.g., temperature, magnetic field, or pressure.
Importantly, NV magnetometry is quantitative, in the sense that it offers self-calibrated, quantitative mapping of the stray magnetic fields in the vicinity of a sample of interest\,\footnote{It is worth noting that NV magnetometry is only strictly quantitative in the regime of weak magnetic fields ($\lesssim10~$mT), where the first order response of the NV ODMR frequencies dominates the Zeeman shift if the NVs spin levels, and transverse magnetic fields only play a negligible role.}.
We provide in the following a brief overview of prominent examples of NV-based static magnetic field imaging for probing a variety of condensed matter targets.
We note that a direct, quantitative determination of physical properties of samples under investigation always needs to be inferred from the measured stray field maps. 
Indeed, careful data-analysis\,\cite{Tetienne2017a} or reverse-propagation\,\cite{Broadway2020a,Dubois2022a} from the measurement plane to the source allows for such quantitative extraction of physical parameters of samples of interest, and for the tracking of their evolution with external quantities such as temperature or magnetic fields.

\subsection{Quantitative Determination of Magnetisation Strength}

The quantitative determination of magnetisation strength of magnetic materials and, e.g., its dependence on temperature of other control parameters, is one of the most fundamental characteristics of any magnetic material. 
However, for samples of mesoscopic scale, very few experimental tools are available to obtain such information. 
NV magnetometry fills this gap and allows for the precise, quantitative determination of magnetization strengths down to the nanoscale. 

Specific examples where NV magnetometry has been put to use in this context include the determination of magnetisation strengths in few layers of magnetically ordered vdW materials, down to the monolayer (Fig. 2a)\,\cite{Thiel2019a, Sun2021a}, and uncompensated surface magnetic moments on magnetoelectric antiferromagnets\,\cite{Appel2019a,Hedrich2021a}.
More conventional thin film magnets have also been examined in this regard\,\cite{Hingant2015a}, but are typically accessible via other experimental approaches as well. 
A notable extension of such studies is the determination of magnetisation strengths in spin-spirals of bulk multiferroics\,\cite{Gross2016a,Finco2022a}. 

The relevance of these results relies on their quantitative character, which, together with the temperature or field dependence of magnetisation, provide key insight into the theoretical understanding of these materials. 
As a further example, NV magnetometry might be highly suited to validate recent theoretical predictions regarding the nature and strength of surface magnetisation in magnetoelectric antiferromagnets\,\cite{Spaldin2021a}.

\begin{figure*}[ht]
	\centering
	\includegraphics[width=0.95\textwidth]{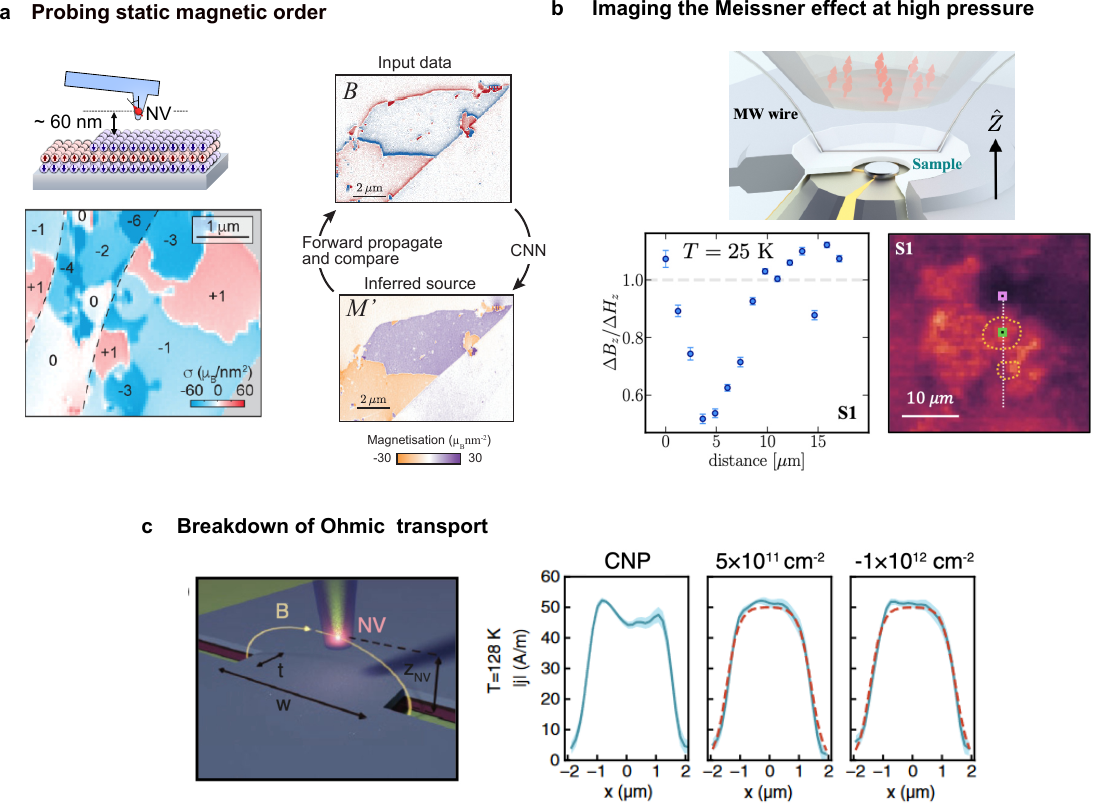}
	\caption{\textbf{Imaging static fields with NV centres. a.} The left panels show an NV centre scanning experiment measuring magnetic domains in a flake of CrI$_3$, revealing regions where the magnetization appears in integer multiples of the monolayer magnetization \cite{Thiel2019a}. To infer the source magnetization from the measured stray field maps (right panels), conventional linear inverse maps may be replaced by more robust convolutional neural networks (CNNs), which require no prior training, and which use physics-informed forward propagation from the magnetisation map $\mathcal{M}'$ to the magnetic field map $\mathcal{B}$ at the measurement plane \cite{Dubois2022a}. \textbf{b.} Locally imaging the Meissner effect in CeH$_9$ at high pressure \cite{Bhattacharyya2023a}. The sample is compressed in a DAC with an NV-containing layer in the upper diamond (top). ODMR measurements of the local magnetic field, taken along the linecut shown in the NV fluorescence image (bottom right), reveal significant inhomogeneities in diamagnetism at the micron scale (bottom left). \textbf{c.} Scanning NV magnetometry of the local electron flow profile through a constriction in graphene \cite{Jenkins2022a}. The local current profile (right, carrier density indicated above each panel) shows characteristic Ohmic peaks at the charge neutrality point (CNP), which breaks down at low temperature away from CNP, indicating the absence of momentum-relaxing scattering as electron-electron scattering becomes more dominant. The bottom left part of panel \textbf{a} is reprinted with permission from \cite{Thiel2019a}. The right part of panel \textbf{a} is adapted with permission from \cite{Dubois2022a}. Panel \textbf{b} is adapted with permission from \cite{Bhattacharyya2023a}. Panel \textbf{c} is adapted with permission from \cite{Jenkins2022a}.} 
	\label{fig:staticfields}
\end{figure*}

\subsection{Magnetisation Distributions}

Quantitative NV magnetometry further enables the disambiguation between different types of domain walls in magnetic materials, such as Bloch and N\'eel domain walls. 
The latter nomenclature describes how the order parameter in a magnetic system evolves across domain walls that separate two domains of homogeneous order parameter -- for Bloch (N\'eel) walls, the evolution is described by a rotation around a vector that is orthogonal to (contained in) the domain wall.  
The identification of the nature of such domain walls is of crucial importance to understanding and characterising the type of exchange interactions that contribute to stabilising a given magnetic order\,\cite{Hubert1998a}, with the Dzyaloshinskii–Moriya interaction being the most prominent example. 
This concept was first put to use in seminal work by Tetienne et al.\,\cite{Tetienne2014b}, who unequivocally demonstrated an experimental distinction between different types of domain walls occurring in thin films of the same magnetic material that was subjected to different interfacial conditions.
Such analyses crucially depend on a proper characterization of the NV orientation and offset distance, the latter being the key parameter of concern.
For reliable, quantitative conclusions, such characterisations are to be conducted in-situ, on the sample under investigation, to avoid systematic errors and concurrent pitfalls in data analysis.

Further, the study of topologically protected magnetic textures, such as magnetic skyrmions\,\cite{Fert2017a} and their winding numbers is directly accessible and enabled by scanning NV magnetometry.
While a range of experiments have demonstrated direct imaging and real-space studies of skyrmions\,\cite{Hrabec2017a,Gross2018a,Rana2020a,Akhtar2019a,Jenkins2019a}, quantitative analysis of such data allows for determining skyrmion winding numbers based on static stray field imaging by scanning NV magnetometry -- a remarkable achievement that has been demonstrated recently\,\cite{Dovzhenko2018a}.
The latter work had a profound impact on skyrmion research, as it was amongst the first to experimentally determine the handedness of individual skyrmions through careful experimentation and data analysis.
A still outstanding challenge is to extend such studies to single-crystalline, skyrmion-hosting materials\,\cite{Seki2012a}, where pristine skyrmion behaviour, including collective modes in skyrmion lattices\,\cite{Soda2023a} -- and possibly even quantum dynamics of skyrmions\,\cite{Psaroudaki2020a} -- could be examined in the future. 

\subsection{Superconducting Order}

By locally probing the magnetic response of a superconductor, it is possible to image superconducting order and various parameters of the superconducting state. 
The magnetic (London) penetration depth $\lambda_L$ determines the lengthscale over which magnetic field lines can enter a superconductor before Meissner screening becomes effective. 
Quantitative determinations of $\lambda_L$, together with its temperature dependence and spatial anisotropy, provide valuable insight into a superconductor's superfluid density and pairing mechanism\,\cite{Prozorov2006a}. Specifically, $\lambda_L$ determines the superfluid density, i.e. the number of electrons that superconduct\,\cite{Luan2010a}.

NV magnetometry has been employed to directly measure the penetration depth, in particular in high-temperature superconductors\,\cite{Thiel2016a,Schlussel2018a,Rohner2019a,Acosta2019a}.
This has for instance been achieved by imaging vortices in type-II superconductors, where a quantitative analysis of vortex stray fields directly yields the London penetration depth\,\cite{Thiel2016a}\footnote{In the cited case, a thin film sample of YBCO was investigated. Since the film under investigation was thin compared to $\lambda_L$, the relevant penetration depth in that case was the ``Pearl length''.}. 
Alternatively, $\lambda_L$ was also determined by direct measurements of Meissner-screening in superconductor microstructures\,\cite{Rohner2019a}, or using wide-field NV magnetometry\,\cite{Joshi2019a}.
This latter approach is notably applicable to type-I superconductors as well.

Such studies of superconductor properties by quantitative NV magnetometry have thus far been applied to reasonably well-understood superconductors (e.g.\ YBCO), deep in the type-II limit of superconductivity.
Future worthwhile extensions include the study of vortices in superconductors at the crossover between type-I and type-II superconductivity, where vortex stray fields are still poorly understood and can only be described by approximative or numerical models\,\cite{Clem1975a,Loudon2015a,Brandt1997a}. 
Such experiments might offer the interesting perspective of simultaneously determining $\lambda_L$ and the superconductor coherence length $\xi$ by direct magnetic imaging. 

NV centres incorporated into a diamond anvil cell (DAC) allow for local magnetic imaging of materials under high, $\sim$ 100 GPa, pressures \cite{Hsieh2019a,Lesik2019a,Yip2019a,Hilberer2023a}. High pressure presents an important knob for tuning microscopic interactions, and has led to the discovery of novel high temperature superconductors \cite{Mao2018a}. Recent work \cite{Bhattacharyya2023a} (Fig. \ref{fig:staticfields}b) used an NV-infused DAC to image the spatial structure of superconducting regions in a hydride superconductor at the micrometer scale, revealing significant inhomogeneities and informing the materials synthesis of superhydrides. Though nontrivial to maintain magnetic sensitivity of NV centres at these high pressures, there are ongoing efforts to extend the pressure range higher \cite{Hilberer2023a,Bhattacharyya2023a}, a frontier for spatially resolved magnetic imaging. 

\subsection{Current Imaging}

Electrical currents produce magnetic fields according to the Biot-Savart law that can be detected and imaged via NV centre magnetometry. Spatial maps of current patterns in a device can distinguish different scattering mechanisms in a material, providing significantly more information than a standard transport measurement that yields just one number - the spatially averaged conductivity. For instance, the path that electrons take in a device depends on the relative strengths of momentum-conserving scattering (e.g.\ electron-electron scattering) and momentum-relaxing scattering (e.g. electron-phonon or impurity scattering). 
Recently, NV magnetometry has been applied to quantitatively probe different transport regimes in a few electronic materials \cite{Ku2020a,Vool2021a,Jenkins2022a}, in particular looking for spatial signatures of current flow in the rare electron hydrodynamic regime, where electron-electron scattering dominates and electrons move akin to fluids.
Several device geometries that produce distinctive flow patterns in the hydrodynamic, ballistic, and Ohmic regime have been proposed and experiments have begun to reveal signatures of hydrodynamics.
However, concerns about the influence of unknown boundary conditions, and the masquerading of ballistic flow and/or small angle scattering as hydrodynamic flow, remain in some cases.
Nevertheless, direct spatial mapping of the current flow provides much more direct evidence of hydrodynamic behavior, and importantly, provides a way to quantitatively probe the importance of electron-electron interactions (Fig.\ 2c), and hence is a good proxy for understanding other open questions in condensed matter, such as the role of electron-electron interactions in the superconductivity of graphene ~\cite{Balents2020}. 
Spatially resolved current imaging can also reveal local current anomalies or irregularities stemming from, e.g.  trivial effects like variations in the electric potential, due to imperfections or impurities, whose presence can greatly affect the interpretation of a standard transport measurement, leading to incorrect conclusions.

\onecolumngrid
\begin{tcolorbox}[colback=boxcolor, colframe=boxcolor,title=\textbf{Box 2: Dynamical correlators from NV noise spectroscopy},coltitle=black,width=\textwidth]

In this box we outline how NV relaxation and decoherence times can be explicitly related to the field noise at the NV centre, and thus to the physical autocorrelation functions of the noise source. To keep the discussion concrete, we will focus on magnetic field noise at the NV centre due to coarse-grained, effectively classical magnetic or current noise in the sample: $H_{\mathrm{NV}} = \omega_0 \mathbf{\hat a \cdot \hat S} + \gamma_\text{e} \mathbf{B}(t) \cdot \mathbf{\hat S}$.
Let us define coordinate axes $a, b, c$ such that ${\bf \hat a}$ points along the NV quantization axis and ${\bf \hat b}, {\bf \hat c}$ are perpendicular to it. For a $T_1$ relaxometry experiment, Fermi's Golden Rule for the spin-1 NV centre \cite{Myers2017a,Zhang2023a} implies 
\begin{align}\label{T1full}
T_1^{-1} = \frac{3}{2} \gamma_\text{e}^2 S_B^{\perp}(\omega_0), \quad S_B^{\perp}(\omega_0) = \int dt \, e^{i \omega_0 t} \langle B_+(t) B_-(0) \rangle, \quad B_\pm = B_b \pm i B_c,
\end{align}
where the NV centre transition frequency $\omega_0$ determines the detected frequency (usually a few GHz).
For dynamical-decoupling based $T_2$ spectroscopy, it is instead more natural to model the NV centre decoherence as a stretched exponential $C(t)=\text{exp}\left[-(t/T_2)^p\right]$, where $p$ is the stretching factor. Then, under dynamical decoupling using e.g.\ N equally spaced pulses with interpulse spacing $\tau$ and total time $T=N\tau$,
\begin{equation}\label{T2full}
T_2^{-p} = \frac{\gamma_\text{e}^2}{\pi T^p}\int_0^\infty d\omega S_B^{\parallel}(\omega) \frac{F_N(\omega\tau)}{\omega^2}, \quad S_B^{\parallel}(\omega_0) = \int dt \, e^{i \omega_0 t} \langle B_a(t) B_a(0) \rangle,
\end{equation}
where the detected noise frequency is set by the pulse rate $\tau^{-1}$ (typically less than $100\,$MHz) through the pulse sequence filter function $F_\text{N}(\omega\tau)=8\sin^4(\omega\tau/4)\sin^2(N \omega\tau/2 )\cos^{-2}(\omega\tau/2)$. For large $N$ this is sharply peaked at $\omega=\pi/\tau$, and it is common to approximate the spectral density as $S_B^\parallel(\pi/\tau)=-\pi \ln[C(T)]/T$ \cite{Romach2015a,Myers2017a,Sangtawesin2019a}. 

\begin{center}
   \includegraphics[width=0.83\textwidth]{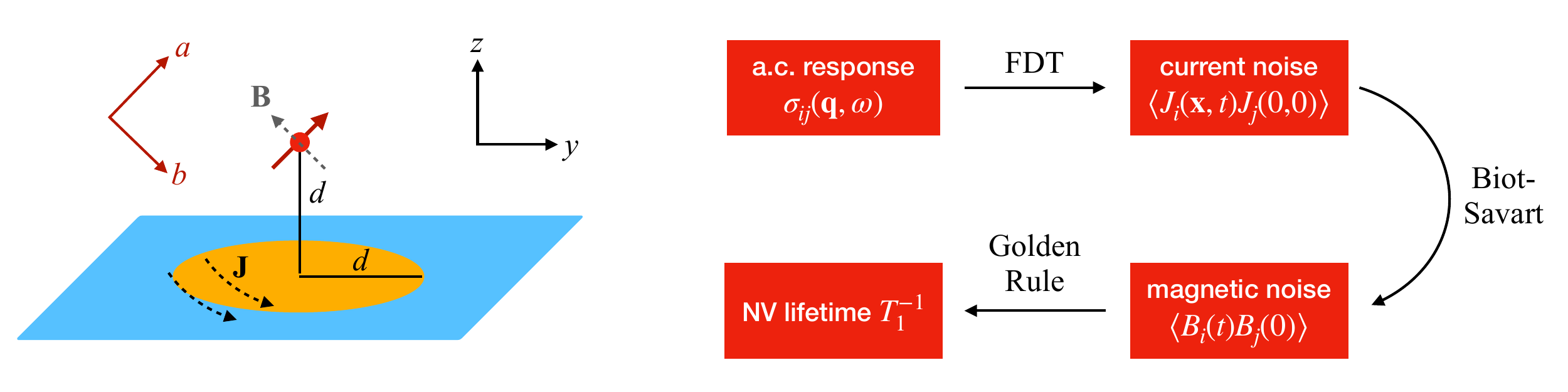}
\end{center}

It remains to calculate the noise spectrum $S_B(\omega)$ at the NV centre from a given source, which will depend on the sample and geometry. As an example, we consider relaxometry in the presence of classical current noise in the sample \cite{Langsjoen2012}. 
The sample is taken to lie in the $xy$ plane and the NV centre is displaced by $d$ along the $z$ axis, aligned along axis ${\bf \hat a}$ (see image above). The current $\mathbf{J}(\mathbf{r},t)$ at a point $\mathbf{r}$ in the sample is a classical field that fluctuates slowly enough that its effect on the NV centre can be treated as instantaneous (this ``quasistatic'' assumption always holds in practice). In this case the NV centre experiences a fluctuating magnetic field that can be related to the current profile through the Biot-Savart law:
\begin{equation}\label{biot1}
\mathbf{B}(t) = \frac{\mu_0}{4\pi} \int d^2 \mathbf{r} \, \frac{\mathbf{J}(\mathbf{r},t) \times ({\bf \hat z} d - {\bf r})}{|r^2 + d^2|^{3/2}}.
\end{equation}
Thus, Eqs.~\ref{T1full} and \ref{biot1} relate the NV centre relaxation time to the current autocorrelation tensor $\langle J_i(\mathbf{x}, t) J_j(\mathbf{0},0) \rangle$ and a spatial smoothing factor from the distance dependence. 
Next, the fluctuation-dissipation theorem (FDT) relates the current autocorrelation function to the current \emph{response} function, $\chi_{ij}(q,\omega)$, which will depend on the conductivity $\sigma_{ij}(q,\omega)$ (although the relation~\ref{biot1} is not precisely $q$-resolved, the $d$-dependence of $T_1$ imposes strong constraints on the $q$-dependence). 
For a 2D conductor at high temperatures such that $\coth(\hbar \omega_0/2k_B T)\approx 2k_BT/(\hbar \omega_0)$, the noise density is \cite{Agarwal2017} 
\begin{equation}\label{spectraldensity}
2S^{x,y}_B(\omega)=S^z_B(\omega)=\frac{\mu_0^2 k_B T }{16 \pi}\varsigma(d,\omega), \quad \varsigma(d,\omega) = \frac{1}{d^2}\int_0^\infty dx\, x \text{e}^{-x}\text{Re}[\sigma^\text{T}(x/2d, \omega)],
\end{equation}
where $\sigma^\text{T}(q,\omega)$ is the $q$-dependent transverse conductivity, which can be probed through the distance-dependence of Eq.~\ref{spectraldensity}. Finally, assuming the NV centre axis ${\bf \hat a}$ makes angle $\theta$ with $z$, the relaxation rate is then $T_1^{-1}=\frac{3}{2}\gamma_e^2 S_B^z(\omega_0)\left(1 + \frac{1}{2}\sin^2(\theta) \right)$ \cite{Ariyaratne2018a}.

A similar procedure may be followed for other noise sources, like magnetic fluctuations \cite{VanDerSar2015a,Huang2023}, or other geometries. For instance, diffusive transport in a conducting film of thickness $a$ and conductivity $\sigma=\text{Re}[\sigma(\omega_0)]\approx\sigma(0)$ produces noise according to Eq.~\ref{spectraldensity} with distance dependence $\varsigma(d,\omega) = (d^{-1}+(d+a)^{-1})\sigma$, which scales as $d^{-1}$ or $d^{-2}$ for thick and thin films, respectively \cite{Kolkowitz2015a,Ariyaratne2018a}. For noise originating from paramagnetic (e.g.\ nuclear) spins, the distance dependence is instead $d^{-6}$ from a single spin \cite{Lovchinsky2016a}, $d^{-4}$ from a 2D layer \cite{Lovchinsky2017a}, or $d^{-3}$ from a 3D half-space \cite{Pham2016}. 

\end{tcolorbox}
\twocolumngrid

\section{Dynamics}

In equilibrium, the response of a material is related to its intrinsic thermal or quantum fluctuations (i.e., the noise it generates) by the fluctuation-dissipation theorem~\cite{kardar2007statistical}: for example, the conductivity at frequency $\omega$ is related to the Fourier transform of the current autocorrelation function, which can be measured using NV magnetometry. In systems that are driven out of equilibrium, fluctuations and response contain distinct information about the state of the system: to characterize the steady state, one needs separate access to both quantities. For example, current noise in nonequilibrium states is sensitive to properties such as the charge of individual carriers~\cite{blanter2000shot}, which equilibrium transport does not give access to. Most solid-state experimental setups used to study noise \cite{Wang2022b,Kobayashi2021a} are restricted to measuring current noise in essentially one-dimensional geometries. NV centres can overcome this restriction and sense local current or magnetic-field fluctuations in two-dimensional systems. They can therefore be used to explore questions about nonequilibrium noise that would not previously have been experimentally accessible. In addition, even in equilibrium, NV centres are unique in having the ability to probe local wavevector-dependent current noise. 

Noise from a nearby material can affect the NV centre either by depolarizing it and decreasing its $T_1$ time or by dephasing it and affecting its $T_2$ time. Accordingly an NV centre can be used as a noise sensor in one of two basic modes, relaxometry ($T_1$) and Ramsey ($T_2$) spectroscopy. In what follows, we outline how an NV centre couples to noise in the underlying sample, then review these two basic spectroscopic protocols, as well as recent extensions that involve measuring correlations among multiple NV centres.

\begin{figure*}[ht]
	\centering
	\includegraphics[width=0.9\textwidth]{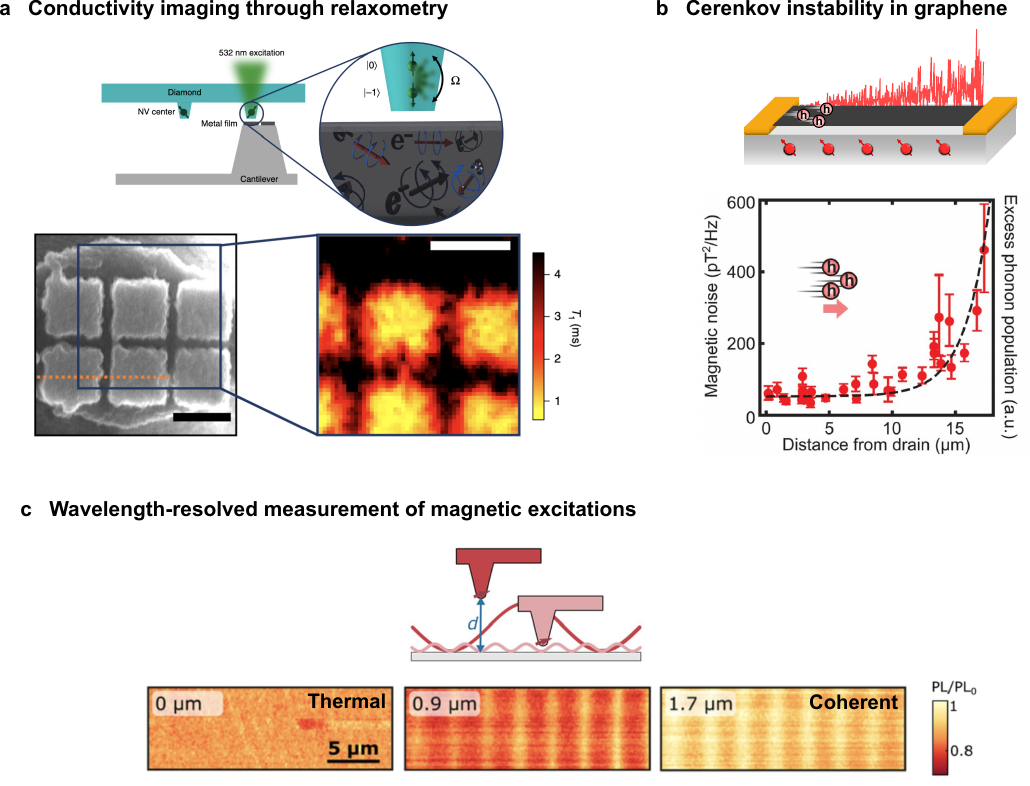}
	\caption{\textbf{Sensing dynamics through relaxometry. a.} Nanoscale conductivity imaging, using NV centre relaxometry induced by thermal electron motion in the metal \cite{Ariyaratne2018a}. Imaging of an Al nanopattern resolves features down to 5 nm (bottom), where the scale bars are 400 nm. \textbf{b.} Electronic noise in graphene exhibiting amplification along the transport direction, detected using proximal NV centres \cite{Andersen2019a}. Spatially resolved noise magnetometry was used to diagnose the anomalous noise as arising from an electron-phonon Cerenkov instability at high current. \textbf{c.} Wavelength-resolved imaging of magnetic excitations in YIG using the distance-dependence of the stray field filter function \cite{Simon2022a}. Spin waves are excited using a microwave stripline, then self-interact through magnon-magnon interactions creating local magnetic noise. Near the sample, short-wavelength incoherent thermal magnons dominate the NV centre relaxation, while further from the stripline, long-wavelength coherent magnons at the NV centre transition frequency drive NV centre Rabi oscillations. Panel \textbf{a} is adapted with permission from \cite{Ariyaratne2018a}. The bottom part of panel \textbf{b} is reprinted with permission from \cite{Andersen2019a}. Panel \textbf{c} is adapted with permission from \cite{Simon2022a}.}
	\label{fig:dynamicsT1}
\end{figure*}

\subsection{What an NV centre senses}

In the subspace of the $0$ and $+1$ magnetic levels of the NV centre, the general Hamiltonian of the NV centre as well as the sample can be written as $\hat H = \gamma_\text{e} B_0 S_z + \gamma_\text{e}\mathbf{\hat B} \cdot \mathbf{\hat S} + \hat H_{\mathrm{sample}}$: the magnetic field that couples to the NV is (in general) operator-valued, so $[\mathbf{\hat B}, \hat H_{\mathrm{sample}}] \neq 0$. While this formulation is general, it is also intractable, and we will make various approximations to arrive at a simpler effective Hamiltonian. 

The first, most natural simplification is to note that the field at the NV centre is determined by the properties of a relatively large patch (of diameter $\geq 10$ nm) of the sample. In most condensed matter contexts, there are many active degrees of freedom in a patch of this size. Even though each degree of freedom is quantum, the average of the current or magnetic field over a patch of this size can be treated as a classical stochastic variable. This allows us to describe the NV as a qubit driven by a time-dependent classical field $\mathbf{B}$, so 
\begin{equation}\label{hnv}
\hat H_{\mathrm{NV}} = \gamma_\text{e}B_0 S_z + \gamma_\text{e}\mathbf{B}(t) \cdot \mathbf{\hat S}. 
\end{equation}
It suffices (i) to relate the statistics of $\mathbf{B}(t)$ to those of the dynamical fluctuations in the underlying sample, and (ii)~to characterize the dynamics of the qubit for $\mathbf{B}(t)$ drawn from such an ensemble. 

A sample can generate a magnetic field at the NV centre either because it has magnetic fluctuations or because it has fluctuating currents $\mathbf{J}(\mathbf{x},t)$ that induce a magnetic field at the NV centre. In either case, there is a linear kernel relating the magnetic field at the NV centre to the relevant fluctuating field in the sample. In general this kernel can be nonlocal in time, and this nonlocality can have significant effects in three-dimensional metals, where current fluctuations have to travel through the metal to reach the NV. However, the situation is simpler in two dimensions, and one can generally make a quasistatic approximation so that the field at the NV centre is linearly related to the current (or, with the obvious modifications, the magnetization) in the sample near the NV centre, by an expression of the form $B_i(t) = \sum_{j = 1}^3 \int d^3\mathbf{x} K_{ij}(\mathbf{x}, t) J_j(\mathbf{x},t)$, where $i, j$ are Cartesian coordinates. 
For quasistatic currents the kernel $K_{ij}$ is given by the Biot-Savart law (Eq.~\ref{biot1}), but more generally it will be some convolution of the current (or magnetization) fluctuations in the sample over a length-scale $d$.

\subsection{$T_1$ spectroscopy}

One of the earliest applications of NV centres to studying transport involved $T_1$ noise spectroscopy of Johnson noise from silver films \cite{Kolkowitz2015a}. The calculation in this case is outlined in Box 2, which shows how one can relate the $T_1$ time of the NV to the autocorrelation tensor of the current in the sample, which in turn can be related (via the fluctuation-dissipation theorem) to the momentum-dependent conductivity.
In the interests of brevity we have sketched the derivation of Box 2 in the simplest case; however, the strategy outlined there can be adapted to any form of current noise under the quasi-static approximation (Eq.~\ref{biot1}). By varying the polarizing field and the height of the NV centre from the sample (e.g., by using an ensemble of NV centres at different heights), one can tune $q$ and $\omega$. In principle, by varying the orientation of the sample with respect to the quantization axis, one can also extract angle-dependent information about the correlations of current noise. 

For diffusive transport in the local approximation, the same formalism may be applied to directly measure conductivity, enabling nanoscale conductivity imaging of inhomogeneous materials as demonstrated by Ariyaratne et al.\ (Fig.\ 3a) \cite{Ariyaratne2018a}. Beyond linear response theory, recent work used spatially resolved $T_1$ relaxometry in conjunction with global transport to study anomalous noise in graphene transport, which was ascribed to an electron-phonon Cerenkov instability (Fig.\ 3b) \cite{Andersen2019a}. The ability to image the noise source using NV centres demonstrated that the noise amplification followed charge carriers rather than current, a critical tool in diagnosing the effect. 

Relaxometry has also been applied to the study of magnetic excitations, for example to study thermally excited magnons in permalloy and yttrium iron garnet (YIG) \cite{VanDerSar2015a,Du2017a}, where the NV relaxation rate is very accurately described by models accounting for the number and spectral density of magnons, as well as the interplay between the magnon k-vector and NV-sample distance. Beyond thermal magnons, related studies explored ferromagnetic resonance in driven magnetic systems, describing the mechanism behind broadband NV spin relaxation and demonstrating coherent interactions between NV centres and spin waves \cite{Wolfe2014a,Wolfe2016a,Wolf2016a,Andrich2017a,Kikuchi2017a,Page2019a}. More recently, NV centres have enabled wavenumber-resolved magnon measurements \cite{Lee-Wong2020,Bertelli2020a, Simon2022a} (Fig.\ 3c), the development of a magnon scattering platform based on real-space imaging of spin waves scattered off a target \cite{Zhou2021a}, and the detection of spin waves generated by nonlinear multi-magnon scattering \cite{MccCullian2020}.

Theoretical studies suggest that $T_1$ spectroscopy can be used to probe other phenomena in correlated electron systems, for instance to analyze Bogoliubov quasiparticles in low dimensional superconductors \cite{Chatterjee2022,Dolgirev2022}. Such experiments can distinguish gapless and fully gapped quasiparticles, enabling identification of different symmetries of the order parameter. Another potential application of this technique is to probe one dimensional systems, such as edge states in topological systems. These studies can elucidate the mechanism of electron scattering, which is still poorly understood \cite{RodriguezNieva2018}. Finally, a promising application of NV relaxometry is to probe spin liquid states, in which magnetic noise should provide signatures of different types of spinon dynamics \cite{Chatterjee2019,Khoo2022}.

\subsection{\label{sec:t2}T$_2$ spectroscopy}

While $T_1$ spectroscopy has the advantage of probing relatively high frequencies (in the GHz range), it has limited bandwidth, and also provides relatively coarse information about the structure of the noise. For probing a wider range of low frequencies, and for addressing more detailed questions such as the character, spectrum, non-gaussianity, etc.\ of the noise, $T_2$ spectroscopy offers a more versatile approach. The basic idea behind $T_2$ spectroscopy is to initialize the NV centre on the equator of the Bloch sphere, and have it precess in the fluctuating field from the sample. Unlike $T_1$ spectroscopy, in this approach one senses the field \emph{along} the quantization axis; for $T_2$ spectroscopy to be useful, the noise at the transition frequency of the NV should be negligible. 

Suppose the NV centre is initialized in the state $\ket{+} \equiv \frac{1}{\sqrt{2}} (\ket{\uparrow} + \ket{\downarrow})$. We choose the quantization axis of the NV to be the $z$ axis of its Bloch sphere, and define $B_\parallel$ to be the component of the local magnetic field along the quantization axis~\footnote{The choice of basis on the Bloch sphere is in general distinct from the real-space coordinate system}. 
After free evolution for a time $t$, the NV centre is in a state $\ket{\psi(t)} = \frac{1}{\sqrt{2}} (\ket{\uparrow} + e^{i \phi_t} \ket{\downarrow})$, where $\phi_t = \gamma_\text{e}\int_0^t dt' \,B_\parallel(t')$. One can interrupt this evolution with a variety of pulse sequences borrowed from the NMR literature \cite{Vandersypen2005}; in this case, the expression for the accumulated phase generalizes to \cite{Klauder1962}
\begin{equation}\label{phi}
\phi_t = \gamma_\text{e} \int_0^t dt' \, s(t') B_\parallel(t'),
\end{equation}
where $s(t')$ is the filter function for a particular pulse sequence. For example, in a standard spin-echo sequence, $s(t') = \mathrm{sign}(t/2 - t')$. At the end of the sequence, one measures the NV centre in the $x$ and $y$ bases; the results can be combined to reconstruct the expectation value $\langle e^{i \phi_t}\rangle$. If the noise has zero mean, and the filter function satisfies $\int_0^t dt' s(t') = 0$, one can use the cumulant expansion to write $\langle e^{i \phi_t}\rangle = \exp(-\langle\phi_t^2\rangle/2 + \ldots)$, where $\ldots$ contains non-gaussian corrections. This relates the dephasing rate of the NV centre to the correlation function
\begin{equation}\label{cphi}
C_\phi(t) = \gamma_\text{e}^2 \int_0^t \int_0^t dt' dt'' s(t') s(t'') \left\langle B_\parallel(t') B_\parallel(t'') \right\rangle.
\end{equation}
Finally, the magnetic field at the NV centre can be related to the current in the sample through Eq.~\eqref{biot1}. The filter function $s(t)$ controls the range of frequencies that contribute to decoherence: e.g., if $s(t)$ is a series of $\pi$ pulses spaced by $\tau$, noise below $\pi/\tau$ will be echoed out and will not contribute to decoherence. While pulse-defined filter function bandwidths are limited by the NV centre driving rates (typically less than 100 MHz), new antenna designs and pulse protocols have extended the range for particular experiments \cite{Savitsky2023a,Casanova2018a,Munuera-Javaloy2023a}.

Unlike $T_1$ spectroscopy, which is primarily sensitive to the power spectrum of the noise, $T_2$ spectroscopy allows one to detect higher-order correlations in the noise. An explicit non-gaussian example, for a qubit driven by telegraph noise, was worked out in \cite{Galperin2006}. Exploring such non-gaussian effects in many-body systems (e.g., near phase transitions) and their signatures in NV magnetometry remains an important task for future research.

Most applications of $T_2$ spectroscopy to date have been in the Gaussian approximation, and make use of a periodic pulse sequence to approximately select a single frequency at a time (i.e.\ assuming that the pulse sequence filter function is nonzero only at a single frequency $f$, see Eq.~\ref{T2full}). This has found use in studying environmental noise spectra to diagnose NV centre dephasing sources, especially near surfaces, where dimensional analysis allows the shape of the coherence decay to report on the nature of the noise bath \cite{Sangtawesin2019a,Dwyer2022a,Davis2023a}. Dynamical decoupling can also be used to selectively couple NV centres to nuclear spins by matching the pulse rate to a multiple of the nuclear Larmor frequency; while most examples of nuclear spin sensing have targeted biological applications \cite{Glenn2018a,Aslam2023a}, an example in condensed matter includes the measurement of nuclear quadrupole resonance in hexagonal boron nitride \cite{Lovchinsky2017a}. Other recent applications of $T_2$ measurements of electronic systems include the observation of increased NV centre coherence times in proximity to a superconductor \cite{Monge2023a} and local superconductor imaging through transverse spin relaxometry leveraging this effect \cite{Luan2015a}. 

Beyond the Gaussian approximation, the shape of the coherence decay can reveal dynamical transitions that provide information about the correlation time of the noise \cite{Dwyer2022a,Davis2023a}, which may be deployed to study dynamics in condensed matter systems (Fig.\ 4a). Such noise spectroscopy can provide insights into phase transitions by measuring the qubit relaxation rate as the system is tuned near its critical point \cite{Machado2023a}. As indicated in Fig.\ 4b, the scaling of the NV centre dephasing $\braket{\phi_t^2}$ with measurement time (as discussed above and in Fig.\ 4a), and in particular, its dependence on temperature T relative to the critical temperature T$_c$ and on the NV-sample distance $d$ relative to the correlation length $\xi$, can report on the critical exponents of the transition as the system is tuned near criticality.

\begin{figure*}[ht]
	\centering
	\includegraphics[width=0.98\textwidth]{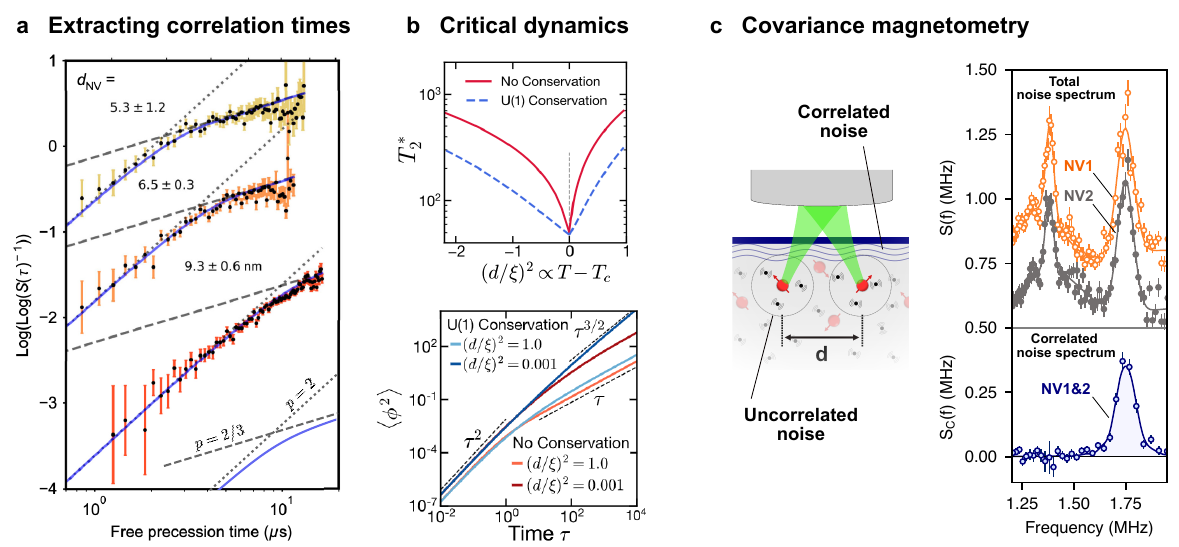}
\caption{\textbf{Ramsey-based noise spectroscopy. a.} The exponential stretching factor (slope of these lines, see Box 2) undergoes a transition at a depth-dependent critical time determined by the correlation time of the noise, from $p=2$ (Gaussian) decay to $p=2/3$, consistent with a reconfiguring 2D noise bath \cite{Dwyer2022a}.
\textbf{b.} Near a phase transition, the dephasing time T$_2^*$ (upper panel) and decoherence dynamics like those shown in \textbf{a} (lower panel) can characterize the transition (e.g.\ diagnose the critical temperature T$_c$ and the presence of order parameter conservation) through their dependence on the system temperature relative to T$_c$ and on the NV-sample distance $d$ relative to the correlation length $\xi$ \cite{Machado2023a}. 
\textbf{c.} The schematic shows two NV centres experience fluctuating magnetic fields that are correlated (as from correlated excitations in a target condensed matter system) or uncorrelated (as from local fluctuating nuclear spins) between the locations of the two NV centres. Whereas two NV centres 1 and 2 may be used as individual spectrometers of their local noise spectra (right, top), covariance spectroscopy reveals which noise sources are correlated and which are uncorrelated between the two NV centres (right, bottom) \cite{Rovny2022a}. Panel \textbf{a} is adapted with permission from \cite{Dwyer2022a}. Panel \textbf{b} is adapted with permission from \cite{Machado2023a}. The right part of panel \textbf{c} is adapted with permission from \cite{Rovny2022a}. }
\label{fig:dynamicsT2}
\end{figure*}

\subsection{Covariance sensing}

The spectroscopic methods described above are spatially resolved in the sense that an NV centre at distance $d$ from the sample will sense noise averaged over a length-scale $\sim d$. However, there are both technical and fundamental limitations to how much spatial information one can extract. At the technical level, moving the NV centre far from the sample makes it couple more weakly, limiting the scope for sensing long-range correlations. More fundamentally, extracting two-point temporal correlation functions from Eq.~\eqref{phi} is only possible if one assumes the noise to be stationary in time; this assumption fails in many nonequilibrium settings. A recently developed technique circumvents these limitations by using multiple NV centres to directly measure correlation functions (Fig.\ 4c) \cite{Rovny2022a}. 

The simplest version of the covariance sensing protocol proceeds as follows. Two NV centres are initialized, at positions $\mathbf{x}_1$ and $\mathbf{x}_2$, and times $t_1$ and $t_2$, along the equator of the Bloch sphere. They precess in a spatio-temporally correlated magnetic field $\mathbf{B}(\mathbf{x},t)$ from the underlying sample. Finally, the phase is mapped to population and both NV centres are measured along the $z$ axis. The classical correlation between these measurement outcomes is related to the correlation between the phase picked up by the two NV centres, which can be reduced (as before) to the noise correlations in the sample. In practice, because of the limited detection efficiency for each measurement, the number of measurements required for a precise estimate of the correlation function is quite large, although it has been mitigated using higher fidelity readout schemes such as spin-to-charge conversion \cite{Shields2015a}. In a proof-of-principle experiment it was shown that a noise sensitivity of approximately $100\,\text{nT}^2/\sqrt{\text{Hz}}$ could be achieved using this approach \cite{Rovny2022a}, improvable with readout methods optimally tailored to detect multiple NV centres. Related recent experiments have used resonant optical readout of NV centres at low temperature to detect correlated electric field dynamics induced by local trapped charges in the diamond \cite{Ji2024a,Delord2024a}.

\section{Conclusions and Perspectives}

In the bulk of this review, we have outlined the ways in which NV centres can be used as nanoscale sensors of both static order and dynamical fluctuations in condensed matter systems. We close with a discussion of some unexplored opportunities that this new technique affords us, as well as some likely near-term technical advances.

NV magnetometry can shed light on the microscopic origins of magnetically ordered condensed matter systems, which exhibit only weak, uncompensated moments. 
A prime example are antiferromagnets, which are typically considered unsuitable for direct magnetic imaging. 
There, NV magnetometry has already found applications, e.g. in magnetoelectrics\,\cite{Appel2019a,Hedrich2021a,Woernle2021a,Meisenheimer2023}, where uncompensated surface moments, collinear with the antiferromagnets' bulk N\'eel vector intrinsically occur\,\cite{Belashchenko2010a} and are measurable by NV magnetometry.
A question of high relevance is how generally applicable NV magnetometry is to the study of antiferromagnetism. 
Indeed, the occurrence of weak, uncompensated moments appears ubiquitous in antiferromagnets. 
Mechanisms such as spin canting, e.g., due to chiral spin-spin interactions\,\cite{Moriya1960a}, or spatially inhomogeneous N\'eel vector spin textures \,\cite{Andreev1996a} have been predicted to lead to such moments and thereby to non-zero stray magnetic fields that may be detectable by NV magnetometry.
Such mechanism would enable generic approaches for, e.g., nanoscale domain imaging in antiferromagnets using static magnetic imaging -- an exciting avenue that remains to be explored in the future.
Further avenues in DC NV magnetometry in condensed matter physics include the nature and existence of time-reversal symmetry breaking in e.g.\ quasi-2D Kagome materials\,\cite{Jiang2023} or in unconventional superconductors such as Sr$_2$RuO$_4$\,\cite{Xia2006a}. Observations of time-reversal symmetry breaking in this presumably nonmagnetic compound could be explained by long-theorized, yet elusive, flux phases\,\cite{Haldane1988a}; local, quantitative, and sensitive experimental probes are necessary to validate the theories. 
For instance, the recent experimental demonstrations of orbital magnetism in Moir\'e systems\,\cite{Tschirhart2021a} imply many potential applications, but very little is known about the underlying origins of this unusual magnetism and its relations to topology and electron correlations.

Beyond imaging magnetic structure, two natural settings in which NV centres can provide fundamentally new information about condensed matter systems are (i)~systems that are strongly spatially inhomogeneous on nanometer length-scales, and (ii)~systems driven out of equilibrium. The former class includes a wide variety of strongly correlated two-dimensional systems, from disordered superconducting films \cite{Sacepe2020} to cuprates \cite{Hayden2023} and Moir\'e materials \cite{He2021}, as well as phenomena like two-dimensional metal to insulator transitions \cite{Abrahams2001,Spivak2004,Ahn2023,Sung2023}. In all of these systems, there is substantial evidence for spatial inhomogeneity, mostly from STM experiments. NV centres offer a complementary probe: instead of measuring the local density of states (as STM does), it measures the local magnetic field at the surface, allowing one to probe finite-frequency fluctuations with very high spatial resolution. The insights from NV sensing have the potential to shed light on many longstanding puzzles involving the role of spatial inhomogeneity, rare-region effects, etc.\ on quantum phases and phase transitions in two dimensions \cite{Nie2014,Vojta2019}. 

The other setting in which NV sensing (especially $T_2$ sensing, covariance magnetometry, and related approaches) can shed new light on many-body physics involves phase transitions and collective phenomena in systems driven out of equilibrium, either transiently (e.g., through a sudden current pulse) or in the presence of a steady-state current drive. Nonequilibrium driving is known to change the critical properties at quantum phase transitions \cite{Mitra2006,Green2006} and also to give rise to phenomena such as light-induced superconductivity \cite{Cavalleri2018}. In these nonequilibrium states, current noise inherently carries information that is distinct from transport; in certain models, one can explicitly show that nonequilibrium noise allows one to distinguish between physical mechanisms that have very similar transport signatures. By shedding light on these mechanisms, nanoscale noise sensing has the potential to discriminate between the many proposed physical mechanisms for unconventional collective behavior in correlated systems.

\begin{acknowledgments}
We wish to acknowledge helpful discussions with Shimon Kolkowitz, Liang Jiang, Ali Yazdani, Pavel Dolgirev, Peter Armitage, Jamir Marino, Atac Imamoglu, Misha Lukin, Hossein Hosseinabadi, Amir Yacoby, and Norman Yao. This work was supported by the National Science Foundation (QuSEC-TAQS OSI 2326767 and Princeton University’s Materials Research Science and Engineering Center DMR-2011750) (N.dL. and S.G.); the Gordon and Betty Moore Foundation (grant DOI 10.37807/gbmf12237) (N.dL.); the Intelligence Community Postdoctoral Research Fellowship Program by the Oak Ridge Institute for Science and Education (ORISE) through an interagency agreement between the US Department of Energy and the Office of the Director of National Intelligence (ODNI) (J.R.); the European Research Council's grant ``QS2DM'' and from the Swiss National Science Foundation through Project $188521$ (P.M.); the Gordon and Betty Moore Foundation’s EPiQS Initiative via Grant GBMF10279 and the DOE QNEXT (A.J.); the DOE Q-NEXT Center (Grant No. DOE 1F-60579) (A.J.); the SNSF project (200021-212899); and the Swiss State Secretariat for Education, Research
and Innovation (contract number UeM019-1)
\end{acknowledgments}

\bibliography{NV_CM_references,NV_CM_referencesPA}

\end{document}